\documentclass[showpacs,reprint,prb,amsmath,amssymb,floatfix, superscriptaddress]{revtex4-1}
\usepackage{bm}					
\usepackage{graphicx}			
\usepackage{xcolor}				
\usepackage{braket}				
\usepackage{mathtools}			
\usepackage{natbib}				

\begin{document}
\title{Fano resonance in a normal metal/ferromagnet-quantum dot-superconductor device}
\author{Lin Li}
\affiliation{Department of physics, Southern University of science and technology of
China, Shenzhen 518005, China}
\author{Zhan Cao}
\affiliation{Center of Interdisciplinary Studies and Key Laboratory for Magnetism and
Magnetic Materials of the Ministry of Education, Lanzhou University, Lanzhou
730000, China}
\author{Hong-Gang Luo}
\affiliation{Center of Interdisciplinary Studies and Key Laboratory for Magnetism and
Magnetic Materials of the Ministry of Education, Lanzhou University, Lanzhou
730000, China}
\affiliation{Beijing Computational Science Research Center, Beijing 100084, China}
\date{\today }
\author{Fu-Chun Zhang}
\affiliation{Department of Physics, Zhejiang University, Hangzhou 310027, China}
\affiliation{Department of Physics and Center of Theoretical and Computational Physics,
The University of Hong Kong, Hong Kong, China}
\author{Wei-Qiang Chen}
\affiliation{Department of physics, Southern University of science and technology of
China, Shenzhen 518005, China}

\begin{abstract}
We investigate theoretically the Andreev transport through a quantum dot strongly coupled with a normal metal/ferromagnet and a superconductor (N/F-QD-S), in which the interplay between the Kondo resonance and the Andreev bound states (ABSs) has not been clearly clarified yet. Here we show that the interference between the Kondo resonance and the ABSs modifies seriously the lineshape of the Kondo resonance, which manifests as a Fano resonance. The ferromagnetic lead with spin-polarization induces an effective field, which leads to splitting both of the Kondo resonance and the ABSs. The spin-polarization together with the magnetic field applied provides an alternative way to tune the lineshape of the Kondo resonances, which is dependent of the relative positions of the Kondo resonance and of the ABSs. These results indicate that the interplay between the Kondo resonance and the ABSs can significantly affect the Andreev transport, which could be tested by experiments.
\end{abstract}

\maketitle

\section{Introduction}

In the past years, the interplay between the Kondo effect and superconductivity has been intensively investigated in hybrid superconductor-nanostructures.
\cite{Rozhkov1999,Clerk2000,Sun2001,Choi2004,Siano2004,Lim2008,Sellier2005,Cleuziou2006,Buizert2007,Karrasch2009,Luitz2012, Pillet2013,Kim2013,Li2014,Rozhkov2000,Buitelaar2002,Buitelaar2003,Graber2004,Meng2009,Koerting2010,Zitko2010,Rodero2011,Lee2012,Kumar2014}
In a superconductor-quantum dot-superconductor (S-QD-S) device, the Josephson current shows an interesting $0$-$\pi$ phase transition at $T_{K}/\Delta\sim1$, where $T_{K}$ is the normal state Kondo temperature, and $\Delta$ is the superconducting gap. At the transition point, the energy gained from the formation of Kondo singlet can exceed the energy gap, which leads to the crossing of Andreev bound states (ABSs). \cite{Rozhkov1999,Clerk2000,Choi2004,Siano2004,Lim2008,Sellier2005,Cleuziou2006,Buizert2007,Karrasch2009,Luitz2012,Pillet2013,Kim2013,Li2014}
Furthermore, if one electrode of the QD is replaced by a normal or ferromagnetic lead, namely, for a normal metal/ferromagnet-quantum dot-superconductor (N/F-QD-S) device, the subgap transport shows much richer features. \cite{Sun2001,Eichler200709,Deacon2010,Hofstetter2010}
For example, the Kondo effect enhancement of Andreev transport has been observed in odd occupation regimes. \cite{Buitelaar2002,Eichler200709}
The coexistence of Kondo resonance and ABSs has been proposed and/or observed in the superconductor-QD devices. \cite{Krawiec2004,Deacon2010,Hofstetter2010,Domanski2008}
However, the interplay between the Kondo resonance and the ABSs in such a device has not been explored in detail.

\begin{figure}[t]
\center{\includegraphics[clip=true,width=\columnwidth]{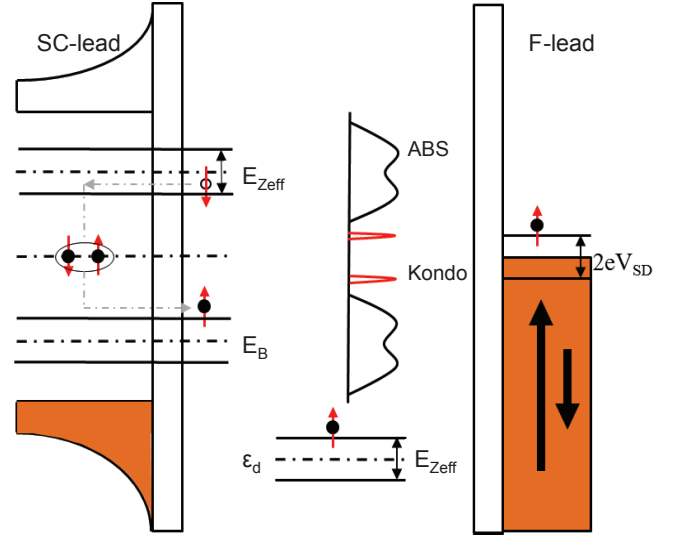}}
\caption{Schematic diagram of the interplay between the Kondo resonance and the Andereev bound states (ABSs) in an F-QD-S device. The spin-polarization induces an effective field $E_{Zeff}$, which removes the degeneracy in dot level $\varepsilon_{d}$ and splits the ABSs and Kondo resonance. When the splitting ABSs and subpeak of Kondo resonance get close to each other, the interference between them modifies the lineshape of the Kondo resonance, which affects the Andreev transport.}
\label{fig1}
\end{figure}

It is well-known that in the N/F-QD-S device, the Andreev levels induced by the proximity of the superconductor will be broadened, whose width is roughly proportional to the coupling strength with the normal or ferromagnetic lead. At the same time, the coupling with the normal or ferromagnetic lead will also lead to a Kondo resonance due to the screening of the spin in the QD if the system is in the Kondo regime. As mentioned above, both these two subgap features have been observed experimentally. \cite{Deacon2010} A natural question arises: is there any interplay between these two subgap features? In this work we explore theoretically the interaction between these two distinguished features and find that the lineshape of the Kondo resonance can be modified seriously by the interference between the Kondo resonance and the ABSs, which can be attributed to Fano resonance. \cite{Fano1961,Miroshnichenko2010}

The experimental observations that the Fano resonance modifies the lineshape of the Kondo resonance have been reported in the experiments about two decades ago, \cite{Li1998,Madhavan1998,Manoharan2000,Jamneala2000} in which the Kondo resonance has been uncovered by scanning tunneling spectroscopy of a magnetic adatom on metal surface. There the asymmetric lineshape of the Kondo resonance observed are resulted from the interference between a symmetric Kondo resonance (discrete channel) and the metal surface states (continuum channel). \cite{Ujsaghy2000} Furthermore, it was found that in the strongly coupling cases, for example,  Ti/Au and Ti/Ag, \cite{Jamneala2000, Nagaoka2002} the broadening impurity level has a significant contribution to the lineshape of the Kondo resonance observed experimentally.\cite{Luo2004} These works indicated that the Fano resonance is popular in such a strongly correlated impurity system (for an overview one can refer to Ref.\,[\onlinecite{Miroshnichenko2010}]).

In this work, we explore the Fano resonance in the N/F-QD-S device. It is shown that the interference between the Kondo resonance and the broadening ABSs has a significant influence on the Andreev transport. In the N-QD-S case, the Kondo resonance peak developed in Andreev transport gradually evolves into an anti-resonance dip structure with increasing the coupling strength of the normal lead and the QD. This can be ascribed to the interference of the Kondo resonance fixed around the Fermi level and the broadening ABSs near the Fermi level. Here the broadening ABSs play a key role in changing the lineshape of the Kondo resonance from a peak to a dip structure. In the F-QD-S case, the situation is more interesting. It is known that the spin-polarization can induce an effective magnetic field $E_{Zeff}$, \cite{Martinek2003,Choi2004b,Swirkowicz2006,Sindel2007,Li2011} which removes the degeneracy of dot level and results in splitting of Kondo resonance and the ABSs. The interference between the sub-peaks of Kondo resonance and splitting ABSs can lead to fine Fano resonance,  as shown schematically in Fig.\,\ref{fig1}. The spin-polarization as well as the magnetic field applied provide a novel way to tune the subgap physics of the Andreev transport in such a device, which can be further tested by future experiments.

The paper is organized as follows. In Sec. \ref{sec2} we introduce the Anderson impurity model and outline the formalism on the Andreev transport.
In Sec. \ref{sec3}, we discuss the numerical results of Andreev transport in the N/F-QD-S device,
and analyze the result obtained by using the Fano resonance picture. Finally, a brief summary is given in Sec. \ref{sec4}.

\section{The model and formalism} \label{sec2}

The N/F-QD-S device can be described by Anderson impurity model \cite{Anderson1961}
\begin{equation}
H=H_{L}+H_{D}+H_{V},  \label{Hamiltonian}
\end{equation}
where
\begin{equation}
H_{L}=\sum\limits_{k\alpha \sigma }\varepsilon _{k\alpha \sigma }c_{k\alpha
\sigma }^{\dagger }c_{k\alpha \sigma }-\Delta \sum\limits_{kS\sigma }\left(
c_{kS\uparrow }^{\dagger }c_{-kS\downarrow }^{\dagger }+H.c.\right)
\label{HL}
\end{equation}
is the Hamiltonian of leads $\alpha =S,N/F$. The second term is only available in the superconducting lead.
The operator $c_{k\alpha\sigma }^{\dagger }$ ($c_{k\alpha\sigma}$) represents the creation (annihilation) of
an electron with the energy $\varepsilon_{k\alpha\sigma }$ in superconductor or normal/ferromagnetic leads, respectively.
\begin{equation}
H_{D}=\sum\limits_{\sigma }\varepsilon _{d\sigma }d_{\sigma }^{\dagger
}d_{\sigma }+\frac{U}{2}\sum\limits_{\sigma }n_{\sigma }n_{\bar{\sigma}}  \label{H-dot}
\end{equation}
is the dot Hamiltonian, in which $\varepsilon _{d\sigma }$ is the dot level, $
d_{\sigma }^{\dagger }$ ($d_{\sigma }$) is the creation (annihilation)
operator of electron in QD, $n_{\sigma } = d_{\sigma }^{\dagger }d_{\sigma }$, and $U$ is the Coulomb repulsion.
\begin{equation}
H_{V}=\sum\limits_{k\alpha \sigma }\left( V_{\alpha}c_{k\alpha \sigma
}^{\dagger }d_{\sigma }+H.c.\right)  \label{H-coupling}
\end{equation}
is the Hamiltonian describing the coupling between the dot and leads with the tunneling amplitude $V_{\alpha}$.
The spin-polarization in ferromagnetic electrode leads to the spin-dependent dot-lead coupling $
\Gamma_{F\uparrow(\downarrow)}=\Gamma_{F}(1\pm P)/2$. $\Gamma _{F
}=\Gamma_{F \uparrow }+\Gamma_{F \downarrow }=\pi\left\vert V_{F
}\right\vert ^{2}\rho _{F }$ is the coupling strength, $\rho _{F}=\rho
_{F\uparrow}+\rho _{F \downarrow}$ is the density of states of the normal/ferromagnetic lead. $P$ is the spin polarization and $P = 0$ denotes a normal electrode. It is known that the ferromagnetic proximity effect induces an effective exchange field on dot level due to the spin-dependent charge
fluctuation. \cite{Swirkowicz2006,Sindel2007,Li2011} This behavior can be treated by the Haldane's scaling theory, \cite{Haldane1978}
and the modification on dot level is
\begin{equation}
\delta\varepsilon_{d\sigma}=\int \frac{d\varepsilon }{\pi }\left\{ \frac{
\Gamma _{F\sigma }\left[ 1-f\left( \varepsilon \right) \right] }{\varepsilon
_{d\sigma }-\varepsilon }+\frac{\Gamma _{F\bar{\sigma}}f\left( \varepsilon
\right) }{\varepsilon -U-\varepsilon _{d\bar{\sigma}}}\right\}.  \label{E_ex}
\end{equation}
The first (second) term denotes charge fluctuation between the single
occupation state and empty (double occupation) state on dot level.\cite{Swirkowicz2006,Sindel2007}

The subgap transports through the device can be derived by Nambu Green's
function
\begin{equation}
\mathbf{G}_{d\sigma }( \varepsilon ) =\left[
\begin{array}{cc}
\langle \langle d_{\sigma };d_{\sigma }^{\dagger }\rangle
\rangle _{\varepsilon } & \langle \langle d_{\sigma };d_{%
\bar{\sigma}}\rangle \rangle _{\varepsilon } \\
\langle \langle d_{\bar{\sigma}}^{\dagger };d_{\sigma }^{\dagger
}\rangle \rangle _{\varepsilon } &\langle \langle d_{%
\bar{\sigma}}^{\dagger };d_{\bar{\sigma}}\rangle \rangle
_{\varepsilon }
\end{array}
\right].  \label{G-Nambu}
\end{equation}
All the components in $\mathbf{G}_{d\sigma }\left(
\varepsilon \right) $ can be calculated by equation of
motion approach. \cite{Lacroix1981, Luo1999, Sun2001,Krawiec2004,Sun2000,Cuevas2001,Baranski2011,Li2014}
In the frame of Hartree-Fock mean field, the dot Green's function reads
\begin{equation}
\mathbf{G}_{d\sigma }^{HF}\left( \varepsilon \right) =\left(
\varepsilon\hat{I} -\tilde{\varepsilon}_{d\sigma }\hat{\sigma}_{z}-\hat{\Sigma}^{0}\left(
\varepsilon \right) -\hat{\Sigma}^{HF}\left( \varepsilon \right) \right)
^{-1},  \label{GF-HF}
\end{equation}
where the non-interacting self-energy $\Sigma _{11}^{0}\left( \varepsilon
\right) =\Sigma _{22}^{0}\left( \varepsilon \right) =-i\left( \Gamma _{F}+
\frac{\theta \left( \left\vert \varepsilon \right\vert -\Delta \right)
\left\vert \varepsilon \right\vert }{\sqrt{\varepsilon ^{2}-\Delta ^{2}}}
\Gamma _{S}\right) ,$ $\Sigma _{12}^{0}\left( \varepsilon \right) =\Sigma
_{21}^{0}\left( \varepsilon \right) =\frac{\Delta }{\sqrt{\Delta
^{2}-\varepsilon ^{2}}}\Gamma _{S}$ with $\Gamma _{S}=\pi \left\vert
V_{S}\right\vert ^{2}\rho _{S}$. $\tilde{\varepsilon}_{d\sigma }=\varepsilon _{d\sigma} + \delta\varepsilon _{d\sigma}$ is the renormalized dot level.
$\delta\varepsilon_{d\sigma}$ plays a role of effective field $E_{Zeff}$ on the dot level.
$\Sigma ^{HF}=\hat{\sigma}_{z}U\langle \hat{n}_{\bar{\sigma}}\rangle +\hat{\sigma}
_{x}\Delta _{d}$ is the interacting self-energy, where $\Delta _{d}=U\langle
d_{\sigma }^{\dagger }d_{\bar{\sigma}}^{\dagger }\rangle $ is the
proximity induced order parameter.\cite{Cuevas2001,Lee2014}

In order to capture the Kondo physics, the higher-order Green's functions should be taken into
account. We truncate the hierarchy of Green's function by Lacroix's
scheme. \cite{Lacroix1981} It qualitatively captures the characters of Kondo
effect even at zero temperature \cite{Kashcheyevs2006} or
particle-hole symmetry. \cite{Qi2009} Although the equation of motion
approach tends to underestimate the Kondo temperature, it can properly
capture the competition between Kondo effect and superconductivity. \cite{Li2014}
After some straightforward calculations, the dot Green's function obtained is
\begin{equation}
\mathbf{G}_{d\sigma }\left( \varepsilon \right) =\left(\varepsilon \hat{I} -
\tilde{\varepsilon}_{d\sigma }\hat{\sigma}_{z}-\hat{\Sigma}^{0}-\hat{\Sigma}
^{K}\right) ^{-1},  \label{GF-Kondo}
\end{equation}
where $\hat{\Sigma}^{K}=\hat{\sigma}_{z}U\left\langle \hat{n}_{\bar{\sigma}
}\right\rangle \frac{\left[ \varepsilon -\tilde{\varepsilon}_{d\sigma}-\Sigma
_{11}^{0}\left( \varepsilon \right) -P_{\sigma }\left( \varepsilon \right)
\right] +Q_{\sigma }\left( \varepsilon \right) /\left\langle \hat{n}_{\bar{
\sigma}}\right\rangle }{\varepsilon -\tilde{\varepsilon}_{d\sigma }-\Sigma
_{11}^{0}\left( \varepsilon \right) -U\left( 1-\left\langle \hat{n}_{\bar{
\sigma}}\right\rangle \right) -P_{\sigma }\left( \varepsilon \right) }+\hat{
\sigma}_{x}\Delta _{d}\frac{\Sigma _{12}^{0}\left( \varepsilon \right) }{
\varepsilon +\tilde{\varepsilon}_{d\bar{\sigma} }-\Sigma _{11}^{0}\left( \varepsilon
\right) }$ is the self-energy containing Kondo correlation, and the notations $P_{\sigma}$ and $Q_{\sigma}$ read
\begin{eqnarray}
&& P_{\sigma }\left( \varepsilon \right) =\left[ \Omega _{1\sigma }\left(
\varepsilon \right) +\Omega _{2\sigma }\left( \varepsilon \right) \right]
+UA_{2\sigma }\left( \varepsilon \right),  \label{P}\\
&& Q_{\sigma }\left( \varepsilon \right) = \left( \varepsilon -\tilde{\varepsilon}_{d\sigma }-U\langle n_{d\bar{\sigma }}\rangle \right) \left[
A_{1\sigma }\left( \varepsilon \right) -A_{2\sigma }\left( \varepsilon
\right) \right] +\langle n_{d\bar{\sigma }}\rangle   \notag \\
&&\hskip 1cm \times \left[ \Omega _{1\sigma }\left( \varepsilon \right) +\Omega
_{2\sigma }\left( \varepsilon \right) \right] -\left[ B_{1\sigma }\left(
\varepsilon \right) +B_{2\sigma }\left( \varepsilon \right) \right],
\label{Q}
\end{eqnarray}
where
\begin{eqnarray}
&&A_{i \sigma }\left( \varepsilon \right) \approx \frac{1}{\pi }\int \frac{
\Gamma _{\sigma }\left( \varepsilon ^{\prime }\right) f\left( \varepsilon
^{\prime }\right) \text{Re}[G_{\sigma }\left( \varepsilon ^{\prime }\right)] }{
\varepsilon -\varepsilon _{i \sigma }}d\varepsilon ^{\prime },  \label{A}\\
&& B_{i \sigma }\left( \varepsilon \right) \approx \frac{1}{\pi }\int \frac{
\Gamma _{\sigma }\left( \varepsilon ^{\prime }\right) f\left( \varepsilon
^{\prime }\right) }{\varepsilon -\varepsilon _{i \sigma }}d\varepsilon
^{\prime },  \label{B} \\
&& \Omega _{i \sigma }\left( \varepsilon \right) =\frac{1}{\pi }\int \frac{
\Gamma _{\sigma }\left( \varepsilon ^{\prime }\right) }{\varepsilon
-\varepsilon _{i \sigma }}d\varepsilon ^{\prime }.  \label{Omega}
\end{eqnarray}
In the above expressions, $\Gamma _{\sigma }\left( \varepsilon \right) =\Gamma _{F\sigma
}+\Gamma _{S}\frac{\left\vert \varepsilon \right\vert \theta \left(
\left\vert \varepsilon \right\vert -\Delta \right) }{2\sqrt{\varepsilon
^{2}-\Delta ^{2}}},$ $\varepsilon _{1\sigma }=\tilde{\varepsilon}_{d\sigma }+\varepsilon^{\prime} -\tilde{\varepsilon}_{d\bar{\sigma }},$
$\varepsilon _{2\sigma}=-\varepsilon^{\prime} +2\tilde{\varepsilon}_{d\bar{\sigma }}+U,i =1,2$. $f\left( \varepsilon \right)$ is the Fermi distribution
function. We can calculate the Green's functions self-consistently and the current through the quantum dot by using the above formulas.

In this work, we study the Andreev transport through N/F-QD-S device,
in which an electron injects from normal/ferromagnetic lead to superconductor through the quantum dot and a hole with oppose spin reflects to
the normal/ferromagnetic lead, then a Cooper pair is created in the superconductor
lead, and vice versa. \cite{Sun2000,Krawiec2004,Baranski2011,Domanski2008}
Therefore, the Andreev current reads
\begin{eqnarray}
I_{A}\left( \varepsilon \right) &=&\frac{2e}{h}\sum\limits_{\sigma }\int
d\varepsilon \Gamma _{F\sigma }\Gamma _{F\bar{\sigma}}\left\vert G_{d\sigma
}^{r}\left( \varepsilon \right) _{12}\right\vert ^{2}\times  \notag \\
&&\left[ f_{F}\left( \varepsilon -eV_{sd}\right) -f_{F}\left( \varepsilon
+eV_{sd}\right) \right],  \label{IA}
\end{eqnarray}
where $f_{F}\left( \varepsilon \pm eV_{sd}\right)$ is the distribution function of electrons in normal metal/ferromagnetic lead, $V_{sd}$ is the bias applied.

\section{Numerical result and discussion} \label{sec3}

\begin{figure}[t]
\center{\includegraphics[clip=true,width=0.8\columnwidth]{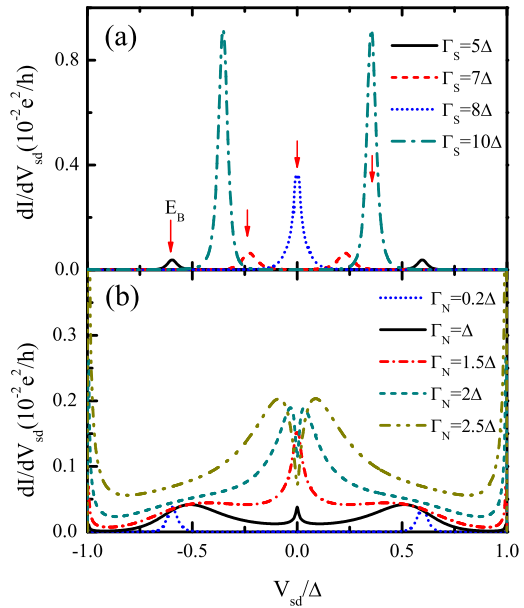}}
\caption{(a) The ABSs dominated by the competition between Kondo
effect and superconductivity in the N-QD-S device with the coupling $
\Gamma_{N}=0.2\Delta$ and $\Gamma_{S}=(5,7,8,10)\Delta$.
(b) The coexistence of Kondo resonance and ABSs with the coupling $\Gamma_{S}=5\Delta$ and $\Gamma_{N}=(0.2,1,1.5,2,2.5)\Delta$.
Other parameters used: the dot level $\protect\varepsilon_{d}=-5\Delta$, the temperature $T=0$, the
magnetic field $E_{Z}=0$, and the half bandwidth $D=20\Delta$.}
\label{fig2}
\end{figure}

In this section, we discuss the numerical results of Andreev transport in the N/F-QD-S device.
In odd occupation regimes, the Kondo resonance appears around the Fermi level of N/F lead due to the screening of local moment.
Here, we pay attention to the interplay between the Kondo resonance and the ABSs, especially, the interference between them.
In the calculations, we assume that the density of states $\rho_{\alpha}=1/2D$ and the temperature $T=0$, $D$ is the half band-width of the N/F lead. In addition, we neglect the influence of applied magnetic field on the leads and set $U\rightarrow\infty$ for simplification.

In Fig.\,\ref{fig2}, we show the Andreev transport in the N-QD-S device, namely, the polarization $P=0$.
When the quantum dot weakly coupled with normal lead ($\Gamma_{N}\ll\Gamma_{S},\varepsilon_{d}$),
the Kondo resonance peak can not be observed in the conductance as shown in Fig.\,\ref{fig2} (a).
In this case, the position of ABSs level $E_{B}$ is determined by the coupling $\Gamma_{S}$.
For $T_{K}\ll\Delta$, the level $E_{B}$ situates away from the Fermi level. In the opposite limit, namely, $T_{K}\gg\Delta$,
the bound states would position above the Fermi level.
The Andreev level $E_{B}$ goes cross the Fermi level about $\Gamma_{S}=8\Delta$ and $T_{K}/\Delta \sim 1$,
which corresponds to the quantum phase transition between magnetic doublet and screened Kondo singlet ground state. \cite{Sellier2005,Kim2013,Li2014,Chang2013}
The width of the ABSs, which associates with its lifetime, only depends on the coupling $\Gamma_{N}$.
In Fig.\,\ref{fig2} (b), we fix the coupling $\Gamma_{S}=5\Delta$ and check the evolution of the Kondo resonance with increasing $\Gamma_N$.
The Kondo resonance peak develops at the Fermi level by increasing the coupling $\Gamma_{N}=\Delta,1.5\Delta$.
Further increasing $\Gamma_N$, the resonance peak evolves into a dip structure, as shown for $\Gamma_{N}=2\Delta,2.5\Delta$.
This change is ascribed to the interference between the Kondo resonance and the significantly broadening ABSs, as discussed later.

\begin{figure}[t]
\center{\includegraphics[clip=true,width=0.8\columnwidth]{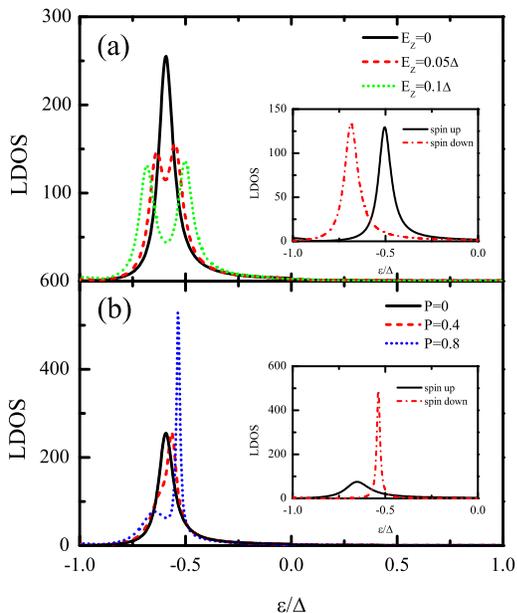}}
\caption{The splitting of the ABSs. (a) The case with different external magnetic field $E_{Z}=0, 0.05\Delta, 0.1\Delta$ and $P=0$.
The inset is the spin-dependent local density of states (LDOS) with $E_{Z}=0.1\Delta$.
(b) The case with different spin-polarization with $P=0, 0.4, 0.8$ and $E_{Z}=0$.
The inset shows the spin-dependent LDOS with $P=0.8$.
Other parameters used are $\varepsilon_{d}=-5\Delta$, $\Gamma_{S}=5\Delta$ and $\Gamma_{N}=0.2\Delta$, and $D=20\Delta$.}
\label{fig3}
\end{figure}

\begin{figure}[t]
\center{\includegraphics[clip=true,width=0.8\columnwidth]{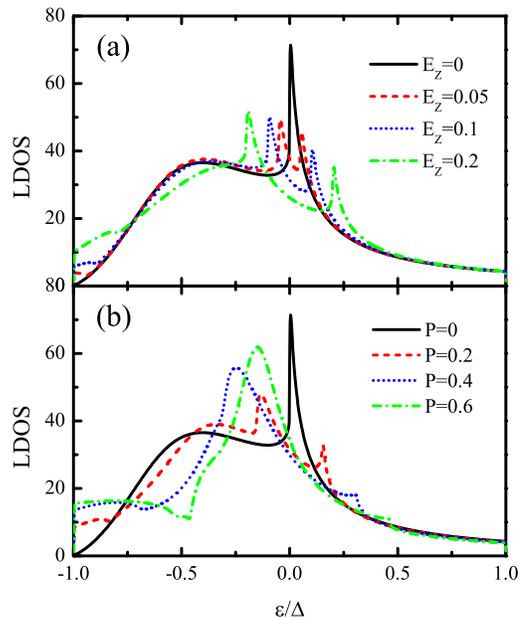}}
\caption{The splitting of the Kondo resonance and the ABSs. (a) The case for different external magnetic field $E_{Z}=0, 0.05\Delta, 0.1\Delta$ and $P=0$.
(b) The case for different spin-polarization with $P=0, 0.2, 0.4$ and $E_{Z}=0$.
Other parameters used are the same as those in Fig. \ref{fig3} except $\Gamma_{N}=1.5\Delta$.}
\label{fig4}
\end{figure}

\begin{figure}[t]
\center{\includegraphics[clip=true,width=0.8\columnwidth]{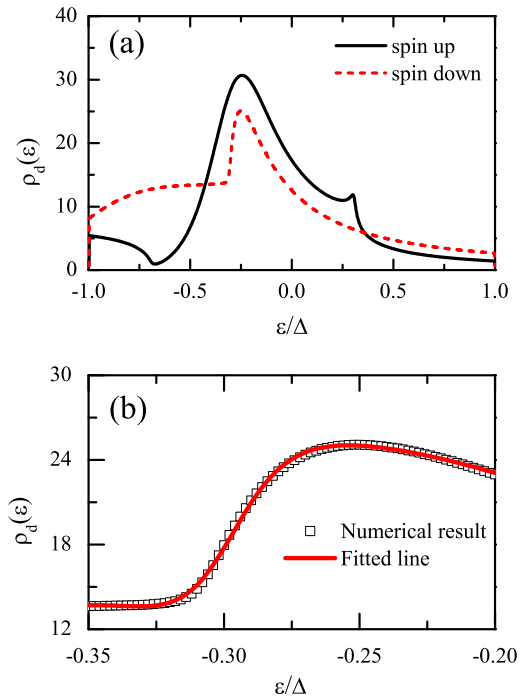}}
\caption{(a) The spin-dependent LDOS with $P=0.4$ and $E_{Z}=0$ as given by the dotted line in Fig.\,\ref{fig4} (b).
(b) The asymmetric structure in the LDOS of spin-down electrons is fitted with the theory of Fano resonance.}
\label{fig5}
\end{figure}

When an external magnetic field is applied and/or the spin polarization in the ferromagnetic lead is present, the Andreev transport displays some interesting splitting behaviors.
In particular, the spin polarization induces a local effective field which removes the spin degeneracy in Kondo effect and ABSs. \cite{Hofstetter2010} In Fig.\,\ref{fig3} (a) and (b), we show the splitting of the ABSs in the presence of Zeeman energy $E_{Z}$ induced by external magnetic field and spin-polarization $P$, respectively. Differently from the splitting induced by external magnetic field (see Fig.\,\ref{fig3} (a)), the ferromagnetic polarization induces an imbalance in the local density of states (LDOS) of spin-up and spin-down electrons, as shown in the inset of Fig.\,\ref{fig3}(b). Due to weak coupling with the dot $\Gamma_N = 0.2\Delta$, the Kondo resonance at the Fermi level does not occur. However, when the quantum dot is strongly coupled with normal or ferromagnetic lead, the Kondo resonance occurs at the Fermi level, as shown in Fig.\,\ref{fig4} (marked by black lines). This Kondo resonance peak becomes splitting when an external magnetic field (see Fig.\,\ref{fig4} (a)) or a ferromagnetic lead (see Fig.\,\ref{fig4} (b)) is applied. Besides the occurrence of the Kondo resonance, another distinguished effect lead by strongly coupling $\Gamma_N = 1.5\Delta$ is the broadening of the ABSs, as shown as a broad peak marked by the black lines in Fig.\,\ref{fig4} (b), which are also splitting under the external magnetic field or the ferromagnetic field. The broadening and splitting of ABSs are found to have a significant influence on the lineshape of the Kondo resonance. For example, in Fig.\,\ref{fig4} (b), for $P = 0.2\Delta,0.4\Delta$ the lineshape of the splitting Kondo resonance peaks is obviously different. For the sub-peak above the Fermi level, the lineshape is a resonance peak, which is similar to that of $P = 0$. On the contrary, the lineshape of the sub-peak below the Fermi level is quite asymmetric(as presented as red line in Fig.\,\ref{fig5} (a)), showing a dip-peak structure, which is reminiscent of the Fano resonance. In the following we argue that the broadening ABSs provides a significant continuum which leads to the asymmetric Kondo sub-peak by interference effect.

It is well-known that the subgap features in the present case involve at most two entities: one is the ABSs and the other is the Kondo resonance if the conditions are met.
Thus, the dot Green's function in the subgap regime can be written by introducing $T$-matrix as
\begin{equation}
G_{d }(\varepsilon )=G_{d }^{0}(\varepsilon )+G_{d }^{0}(\varepsilon
)T_{d}(\varepsilon )G_{d }^{0}(\varepsilon ),  \label{G_d}
\end{equation}
where $G_{d }^{0}(\varepsilon )$ is the dot Green's function in the subgap, which is nothing but the ABS described approximately by a simple Lorentzian resonance
\begin{equation}
G_{d }^{0}(\varepsilon ) = \frac{Z_{B}}{\varepsilon -E_{B}+i\Gamma _{B}},  \label{Gd_0}
\end{equation}
where $E_B, \Gamma_B$, and $Z_B$ denote the position, the width, and the weight of the ABSs.  It is known that $\Gamma _{B} \propto \Gamma_{N/F}$. The $T$-matrix $T_{d}(\varepsilon )$ is
mainly composed by Kondo resonance in the Kondo regime
\begin{equation}
T_{d}(\varepsilon )\approx \frac{\Gamma _{K}}{\pi \rho _{d }^{0}(\varepsilon
)}\frac{1}{\varepsilon -\varepsilon_{K}+i\Gamma _{K}},  \label{Matrix_T}
\end{equation}
which is also approximately described by a Lorentzian line-shape.\cite{Luo2004} $\Gamma_{K}=T_{K}$ is the half-width of Kondo resonance, $\varepsilon _{K}$
is its position.

From Eqs.\,(\ref{G_d})-(\ref{Matrix_T}), the local density of states (LDOS) of the dot in the subgap $\rho_d(\varepsilon) = -\frac1\pi\text{Im}G_d(\varepsilon)$ reads
\begin{eqnarray}
&& \rho _{d }(\varepsilon ) = -\frac{1}{\pi }\text{Im}\left[ G_{d
}(\varepsilon )\right]  \notag \\
&& \hspace{1cm} =\rho _{d}^{0}(\varepsilon )-\pi \left[ \rho _{d }^{0}(\varepsilon )\right]
^{2}\times  \notag \\
&&\hspace{1.3cm} \left[ \left( q_{d}^{2}(\varepsilon )-1\right) \text{Im}T_{d}(\varepsilon
)-2q_{d}(\varepsilon )\text{Re}T_{d}(\varepsilon )\right],  \label{Rho_d}
\end{eqnarray}
where $\rho _{d }^{0}(\varepsilon )=-\frac{1}{\pi }$Im$\left[ G_{d}^{0}(\varepsilon )\right]$ and $q_{d}(\varepsilon )=-$Re$G_{d }^{0}(\varepsilon)/$
Im$G_{d }^{0}(\varepsilon )$ is the asymmetry factor.  Substituting $T_{d}(\varepsilon )$ into Eq.\,(\ref{Rho_d}),
one can obtain a Fano-like formula for the LDOS in the subgap regime
\begin{equation}
\rho _{d}(\varepsilon) = \rho _{d }^{0}(\varepsilon )\frac{(x + q_d(\varepsilon ))^2}{x^2 + 1},  \label{Rho_d2}
\end{equation}
with $x=\left( \varepsilon -\varepsilon _{K}\right) /\Gamma_{K}$, as given in Ref.\,\onlinecite{Luo2004}.
The density of states $\rho^{0}_{d}(\varepsilon )$ can be obtained from Eq.\,(\ref{Gd_0}),
\begin{equation}
\rho^{0}_{d}(\varepsilon )= \frac{1}{\pi }\frac{Z_{B}\Gamma _{B}}{\left(
\varepsilon -E_{B}\right) ^{2}+\Gamma _{B}^{2}}.  \label{Rho_da}
\end{equation}
Eq.\,(\ref{Rho_d2}) can simply capture the in-gap LDOS and the characteristics in Andreev transport,
which can be recognized as the coherent superposition of ABSs and Kondo resonance.

\begin{figure}[t]
\center{\includegraphics[clip=true,width=0.8\columnwidth]{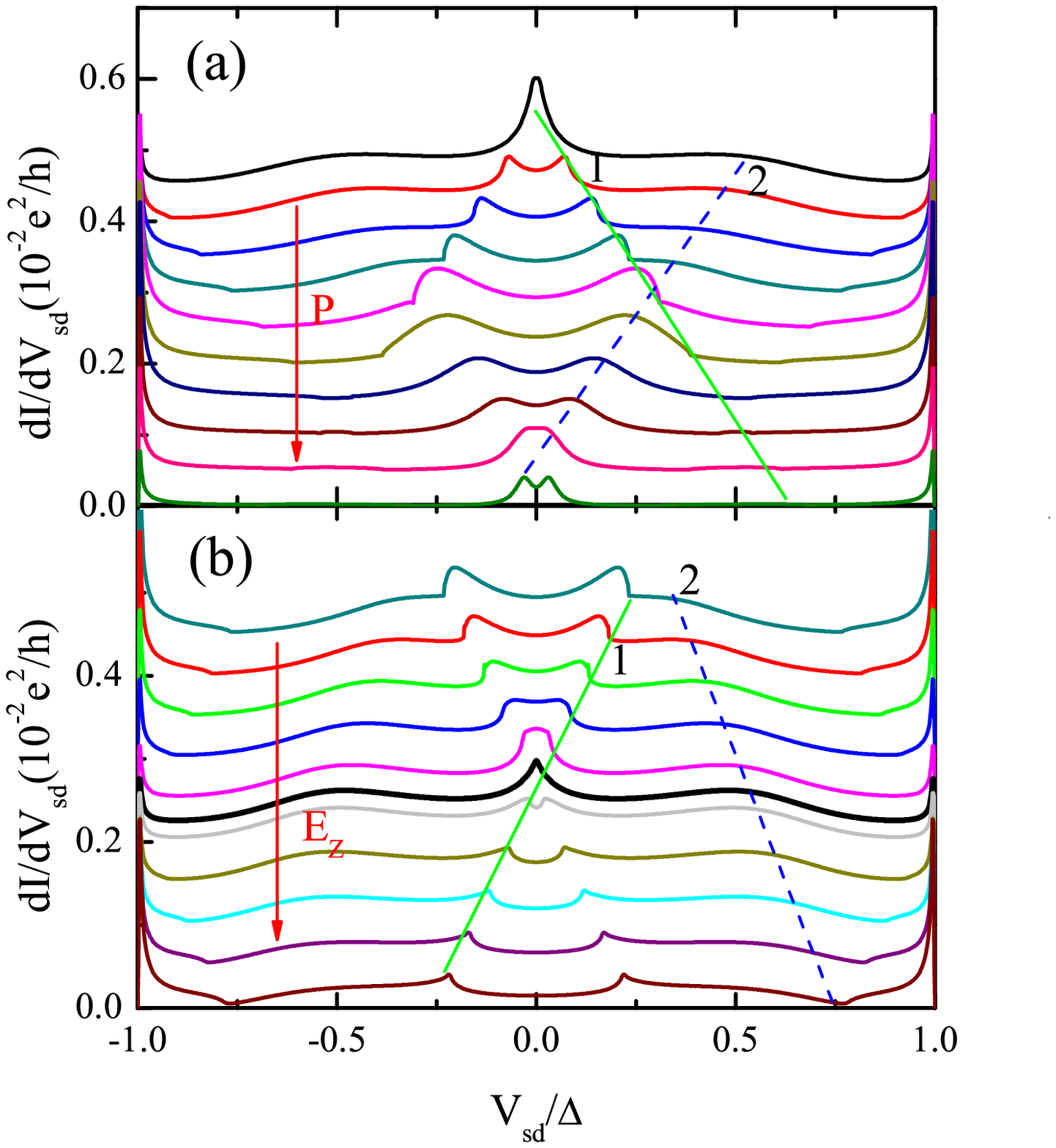}}
\caption{(a) The evolution of ABSs and the splitting of the Kondo resonance due to the spin
polarization of lead, the polarization $P$ increases from $P=0$ (the
top) to $P=0.9$ (the bottom) with the step $0.1$, and $E_{Z}=0$. (b) The
compensation effect of the ABSs and the Kondo splitting by external magnetic fields with $P=0.3$. The Zeeman energy varying from the top $E_{Z}=0$ to the
bottom $E_{Z}=0.225\Delta$ with a step $0.025\Delta$. The compensation field is
$E^{c}_{Z}=0.135\Delta$, as shown the bold dark curve. Other parameters used: the dot level $
\protect\varepsilon_{d}=-5\Delta$, the coupling $\Gamma_{S}=5\Delta$ and $
\Gamma_{N}=1.5\Delta$, and the half bandwidth $D=20\Delta$.}
\label{fig6}
\end{figure}

In the absence of Kondo resonance ($\Gamma_{N}\ll|\varepsilon_{d\sigma}|$), the ABSs always keep the Lorentzian line shape even the factor $|q_{d}|\rightarrow\infty$.
By increasing the coupling $\Gamma_{N}$, the Kondo resonance peak appears and evolves into a Fano-dip structure for $\Gamma_{N}\sim|\varepsilon_{d\sigma}|$ as shown in Fig.\,\ref{fig2} (b).
In this case, the interference between the Kondo resonance and the significant broadening ABSs distorts the line shape of Kondo resonance.
In the presence of magnetic field or spin-polarization, the interference between the subpeak of Kondo resonance and splitting ABSs also leads to Fano asymmetric structures as shown in Fig.\,\ref{fig4}.
In Fig.\,\ref{fig5} (a), we plot the spin-dependent LDOS with $P=0.4$.
The Zeeman splitting of the square-root singularity situates symmetrically around the energy gap. This phenomenon has been reported in the previous experimental and theoretical investigations. \cite{Domanski2008,Meservey1970}
Interestingly, the Kondo resonance splits into a resonant sub-peak and a Fano-type asymmetric structure for spin-up and spin-down electrons, respectively.
In Fig.\,\ref{fig5} (b), we fit the asymmetric structure in the LDOS of spin-down electrons by the qualitative theory of Fano resonance[see Eq.\,(\ref{Rho_d2})] by using the following parameters: $q_{d}=1.15$, $T_{K}=0.031\Delta$, $E_{B}=-0.2\Delta$, $\varepsilon_{K}=-0.3\Delta$, $Z_{B}=0.04$, and $\Gamma_{B}=0.12\Delta$, and the level $E_{B}=E^{0}_{B}+E_{Zeff}$  with $E^{0}_{B}=-0.5\Delta$ and $E_{Zeff}=0.3\Delta$,
where $E^{0}_{B}$ is the bare level of ABS in the absence of spin polarization, and $E_{Zeff}$ is the Zeeman energy induced by the effective field.

\begin{figure}[t]
\center{\includegraphics[clip=true,width=0.8\columnwidth]{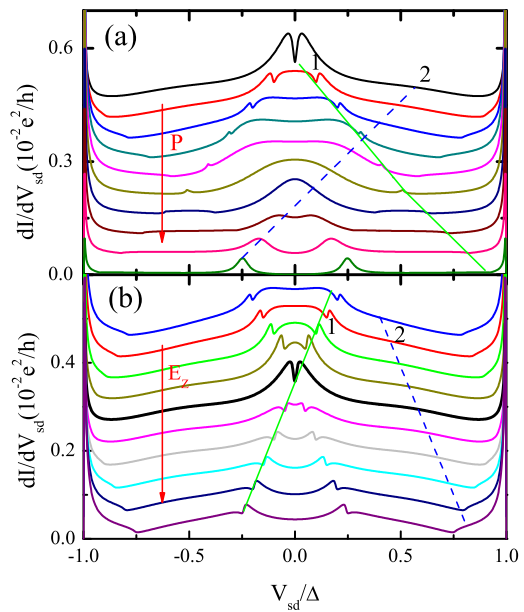}}
\caption{The suppression of ABS and the splitting of Kondo dip due to the spin polarization and its compensation effect by external magnetic field.
(a) The polarization $P$ increases from $P=0$ (the top) to $P=0.9$ (the bottom) with the step $0.1$ and the Zeeman energy $E_{Z}=0$.
(b) The compensation of ABS and the Kondo splitting by external magnetic field varying from the top $E_{Z}=0$ to
the bottom $E_{Z}=0.225\Delta$ with the step $0.025\Delta$. For the polarization $P=0.2$, the compensation field is $E^{c}_{Z}=0.1\Delta$, as shown the bold dark curve.
Other parameters used : the dot level $\protect\varepsilon_{d}=-5\Delta$, the coupling $\Gamma_{S}=5\Delta$ and $
\Gamma_{N}=2\Delta$, and the half bandwidth $D=20\Delta$.}
\label{fig7}
\end{figure}

In Fig.\,\ref{fig6} and  Fig.\,\ref{fig7}, we show the evolution of Kondo resonance peak ($\Gamma_{N}=1.5\Delta$) 
and Fano dip ($\Gamma_{N}=2\Delta$) structures in Andreev transport spectrum of F-QD-S device.
When the ferromagnetic lead is weakly coupled with quantum dot, the spin polarization induced effective field $E_{Zeff}$ is too small to be observed in ABSs.
Therefore, we discuss the case that the dot is strongly coupled both with ferromagnetic and superconductor leads.
From Fig.\,\ref{fig6} (a), we can see that the ABSs are split and suppressed by spin polarization as indicated by the blue dashed line $2$.
In addition, the Kondo resonance is also split into two side-peaks by the induced effective field. \cite{Martinek2003,Choi2004b,Swirkowicz2006,Sindel2007,Li2011}
By increasing $P$, the lineshape of the side-peaks evolves from a peak to a peak-dip, and even to a dip. Eventually, the Kondo effect is completely suppressed by the large polarization $P$, as indicated by the green dotted line $1$.
As shown above, the Fano resonance originates from the interference between the side-peaks of Kondo resonance and the splitting ABSs.
At the bias $V_{sd}=\pm\Delta$, the Kondo resonance at the Fermi level of ferromagnetic lead also significantly enhances the Andreev transport as shown by the green dotted lines $1'$.
In Fig.\,\ref{fig6} (b), we show the compensation effect in Andreev transport,
since external applied field can counteract the influence of effective field on electron spin.
All the splitting structure in ABSs can be compensated by applied magnetic field.
For the polarization $P=0.3$, the corresponding compensated Zeeman energy is $E^{c}_{Z}=0.135\Delta$ as shown the bold dark curve in Fig.\,\ref{fig6} (b).
By increasing the external magnetic field, the asymmetric structures firstly merge into a Kondo resonance peak.
For $E_{Z}>E^{c}_{Z}$, the Kondo resonance peak splits again into two sub-peaks, which become more and more symmetric,
because the Fano interference is suppressed by increasing the distance between the Kondo resonance and ABSs through the Zeeman splitting $E_{Z}$.

In Fig.\,\ref{fig7} (a), we show the influence of effective field on anti-resonance dip structure in Andreev transport.
In the absence of polarization, the anti-resonance dip originates from the interference between Kondo resonance and ABSs.
In the presence of polarization, the interference between the sub-peaks of Kondo resonance and splitting ABSs can be tuned by increasing the distance between them.
As indicated by the green solid line $1$, the resonance dip splits into two side-dip structures, which evolve into resonance peaks and then are completely suppressed by increasing the polarization $P$.
The ABSs are also split by the effective field and shows crossing behavior.
In Fig.\,\ref{fig7} (b), we show the splitting of ABSs and Kondo resonance can be compensated by applied magnetic field.
For the polarization $P=0.2$, the splitting in ABSs can be compensated by the applied magnetic field with $E^{c}_{Z}=0.1\Delta$.
The anti-resonance dips display crossing behavior and evolve into asymmetric structures by increasing the magnetic field.
In Fig.\,\ref{fig6} (b) and Fig.\,\ref{fig7} (b), the Fano asymmetric structures at finite-bias are pronounced when the splitting ABSs gets close to the sub-peak of Kondo resonance.
The line shape is determined by the Fano factor $q_{d}$.
Oppositely, the Kondo resonance recovers the peak structure when it is away from the ABSs.
Because the dot level can not be significantly broadened by applied magnetic field,
the Fano effect can only attributes to the interference between the ABSs and sub-peak of Kondo resonance.

\section{Summary} \label{sec4}

In conclusion, we have studied the Fano resonance in Andreev transport through the N/F-QD-S device.
In N-QD-S device, the interference between the Kondo effect and the broadening ABSs leads to an anti-resonance structure at zero-bias in Andreev transport.
In the presence of spin-polarization, the ferromagnetic lead induces an effective magnetic field, which splits the Kondo resonance and ABSs.
All the splitting structures can be compensated by external applied magnetic field.
When the sub-peaks of Kondo resonance and splitting ABSs get close to each other,
the interference between them results in Fano-type asymmetric structures at finite-bias.
The Fano resonance in Andreev transport is expected to be observed in future experiments.

L. Li gratefully acknowledge useful discussions with Dr.\,Hua Chen. This work is supported by NSFC (Grants Nos. 11174115, 11325417, 11204186, 11274269), PCSIRT (Grant No. IRT1251) of China,
and Natural Science Foundation of Guangdong Province of China 2014A030310137.

\end{document}